\documentclass[sigconf,authorversion,nonacm]{acmart}
\AtBeginDocument{%
  }

\usepackage{graphicx} 
\usepackage{xcolor}
\usepackage{hyperref}
\usepackage{amsfonts}
\usepackage{tabularx}
\bibpunct[, ]{(}{)}{;}{a}{}{,}
\usepackage{amsmath}
\graphicspath{{./figures/}}

\usepackage{soul} 
\newcommand{\stkout}[1]{\ifmmode\text{\sout{\ensuremath{#1}}}\else\sout{#1}\fi} 

\date{September 7, 2026}
\begin{document} 

\title{Evaluating Brand Retrieval and Ranking in Large Language Model Recommendations}


\author{Edward Malthouse}
  \orcid{0000-0001-7077-0172}
  \email{ecm@northwestern.edu}
\affiliation{%
  \institution{Northwestern University}
  \streetaddress{1845 Sheridan Road}
  \city{Evanston} \state{IL}
  \country{USA}
}
\author{Kun-Yu Lee}
\email{Kun-YuLee2027@u.northwestern.edu}
\affiliation{\institution{Northwestern University}\city{Evanston} \state{IL}\country{USA}}

\author{Jing Yang}
\email{jyang15@bu.edu}
\affiliation{\institution{Boston University}\city{Boston} \state{MA}\country{USA}}
\author{Sanchary Pal}
\email{sanchary.pal@northwestern.edu}
\affiliation{\institution{Northwestern University}\city{Evanston} \state{IL}\country{USA}}
\author{Xueyan Feng}
\email{XueyanFeng2026@u.northwestern.edu}
\affiliation{\institution{Northwestern University}\city{Evanston} \state{IL}\country{USA}}

\renewcommand{\shortauthors}{Malthouse et al.}

\begin{abstract}
Large language models (LLMs) are increasingly used for product recommendation, but evaluating their recommendations presents challenges that differ from conventional information retrieval and recommender systems. LLMs can generate recommendations without an explicit candidate set, and repeated responses to the same query can produce different brands and rankings. We introduce a framework for evaluating open-ended LLM brand recommendations that defines the competitive set independently of model outputs and estimates recommendation prevalence and prominence through repeated sampling. We operationalize these constructs using Brand Recommendation Probability (BRP@$k$) and Mean Reciprocal Rank (MRR@$k$), and apply the framework to six LLMs across five product categories. Category-only queries reveal substantial omission of established brands and limited evidence that recommendation prominence follows conventional brand popularity. Instead, prominence is associated with broader marketplace-visibility signals, particularly search interest and online brand conversation. Needs-based queries show that contextualizing users' goals and constraints changes which brands are retrieved, while diagnostic positioning probes demonstrate that brands omitted from ordinary recommendations can remain conditionally retrievable when distinctive cues are supplied. These findings highlight the need to evaluate LLM recommendation as a stochastic retrieval-and-ranking process rather than from individual generated lists. We provide open-source software and data to support reproducible evaluation of LLM-generated brand recommendations.
\end{abstract}


\maketitle

\section{Introduction}

Large language models (LLMs) are increasingly consulted for product information and advice, changing how consumers encounter and evaluate brands. Rather than beginning a decision journey with a search engine or a brand's website, consumers can ask an LLM questions such as ``What is the best detergent for sensitive skin?'' or ``Which laptop should I buy for graphic design work?'' Such requests can be contextual and intent-rich, expressed in natural language that conveys consumers' goals, constraints, preferences, and intended uses. The LLM interprets these needs and recommends a small set of alternatives. In doing so, it becomes an intermediary between consumer needs and brands, potentially influencing which brands enter consideration and the order in which consumers encounter them. A brand omitted from an LLM's recommendations may never be considered, even when it fits the consumer's needs.

This new form of gatekeeping creates a measurement problem for marketers. The set of brands an LLM may recommend is neither fixed nor directly observable, and repeated requests can produce different recommendations and rankings. Brand managers therefore need methods for determining whether their brands are recommended, how frequently and prominently they appear, which competitors appear instead, and how recommendations change when consumers provide information about their needs. Industry and academic analysts face a related problem: understanding how LLM recommendations represent the broader competitive landscape, including brands that are rarely or never recommended.

Two issues are particularly important. First, LLM recommendations may favor a relatively small subset of brands. The marketing and recommender-systems literatures have documented popularity bias, whereby already-popular items can receive disproportionate exposure while long-tail alternatives receive less \cite{abdollahpouri2021user,fleder2009blockbuster}. Whether LLM recommendations similarly reflect brand popularity is unclear; they may instead be associated with other factors, such as marketplace visibility or brand positioning. Second, consumers can communicate their goals, preferences, and constraints directly to LLMs. Such needs-based prompts provide information that may help LLMs match consumer needs with brands, potentially bringing otherwise omitted brands into the recommendation set. This possibility is especially important for firms that invest in differentiated positions intended to associate their brands with particular consumer needs, use cases, and value propositions.

This article introduces a framework for measuring, explaining, and diagnosing LLM-generated brand recommendations. We first define the competitive set independently of the LLM, making it possible to identify not only recommended brands but also omitted ones. We then define \emph{Brand Recommendation Probability} (BRP) to measure recommendation prevalence and adapt \emph{Mean Reciprocal Rank} (MRR) to measure recommendation prominence. Using repeated recommendations from multiple LLMs across five product categories, we examine category-only recommendations and explore marketplace factors associated with recommendation prominence. We then use needs-based prompts to examine whether brands surface when consumers express relevant goals and constraints, followed by diagnostic positioning probes to assess whether omitted brands can be retrieved when supplied with distinctive cues consistent with their intended positioning. We provide an open-source implementation of the framework to facilitate reproducible research and practical brand audits.

\section{Conceptual Background and Framework}

\subsection{LLMs as Choice Architects}

Choice architecture describes how the environment in which alternatives
are presented, including which options appear, how many appear, and how
they are ordered, shapes choice \cite{thaler2009nudge,johnson2012beyond}.
Marketing has long involved such architectures. Retailers
determine which brands appear on physical shelves and where they are
placed; search engines rank pages and sell sponsored positions; and
recommender systems select and rank alternatives from platform
inventories. Each environment has identifiable mechanisms through which
brands gain exposure and established metrics for monitoring that
exposure. Table~\ref{tab:choice-environments} summarizes these environments and contrasts them with
LLM-mediated recommendations.

\begin{table*}[t]
\caption{Comparison of Brand Choice Environments}
\label{tab:choice-environments}
\centering
\small
\begin{tabularx}{\textwidth}{
    @{}
    >{\raggedright\arraybackslash}p{1.7cm}
    >{\raggedright\arraybackslash}p{2.4cm}
    >{\raggedright\arraybackslash}X
    >{\raggedright\arraybackslash}p{3.1cm}
    >{\raggedright\arraybackslash}p{2.6cm}
    >{\raggedright\arraybackslash}p{2.7cm}
    @{}
}
\toprule
\textbf{Choice environment} &
\textbf{Architect} &
\textbf{How the option set is formed} &
\textbf{Brand's levers} &
\textbf{Established monitoring metrics} &
\textbf{Source} \\
\midrule
Physical shelf &
Retailer (category manager) &
Stocking and placement decisions made from a known catalog; planograms
and slotting allowances &
Distribution, trade promotion, slotting fees, packaging &
Share of shelf; distribution coverage &
\cite{dreze1994shelf, chandon2009does} \\
\addlinespace
Search engine &
Ranking algorithm and position auction &
Ranks pages that already exist by relevance signals and bids &
Search engine optimization; paid search bids &
Rankings; share of search; CTR; conversion &
\cite{ilfeld2002generating, choi2012predicting, hu2014decomposing} \\
\addlinespace
Recommender system &
Platform RS designers &
Selects and ranks alternatives from an explicit candidate set or
platform inventory &
Assortment on the platform; sponsored placements &
Precision; recall; MRR; NDCG; CTR; conversion &
\cite{schafer1999recommender,fleder2009blockbuster}
\\
\addlinespace
LLM recommendation &
Model/provider; no direct list curator &
Generates brand names from statistical patterns in training text &
Largely unknown; presence and framing in the text models learn from &
BRP; MRR &
Proposed here \\
\bottomrule
\end{tabularx}
\end{table*}

LLMs differ from these environments in an important respect: rather than
selecting alternatives from a fixed catalog or inventory, they can
generate a set of brand recommendations in response to a
consumer\textquotesingle s prompt. We therefore describe LLMs as
emergent choice architects: they construct a small set of alternatives
at the moment of recommendation, even though no category manager or
other actor has explicitly selected the brands that will appear. For
marketers, this makes brand exposure less transparent. There is no shelf
allocation, search ranking, or platform inventory to inspect; instead,
the brands that surface must be observed in LLM responses, with the
model itself acting as gatekeeper to that visibility \cite{kuenzler2026communication}.

\subsection{From Consideration Sets to Generated Sets}

Consumer research has long recognized that choice proceeds in stages.
From the universe of available brands, consumers form a smaller
consideration set from which they make their final choice \cite{shocker1991consideration, howard1969theory}.
Consideration sets tend to be small
because retrieving and evaluating alternatives is costly \cite{hauser1990evaluation}. Which brands enter the set depends partly on
retrieval: situational cues can activate different brands and usage
associations in consumers\textquotesingle{} memories \cite{nedungadi1990recall, ratneshwar1991substitution}.
Being considered is therefore a
prerequisite for being chosen.

Decision aids can partially externalize this process. Interactive
shopping agents reduce the alternatives consumers consider, while the
ordering of those alternatives concentrates attention near the top of
the list \cite{haubl2000consumer,diehl2003smart,diehl2005two}. LLMs extend this externalization. Consumers can describe a need,
goal, or problem in natural language and ask the LLM what they should
buy. The model then returns a small set of brands on the
consumer's behalf. We call this a generated set: a
compact set of alternatives constructed by the LLM in response to the
consumer's prompt. Unlike an assortment or catalog-based
top-$k$ list, membership and ordering in a generated set may vary
across repeated queries.

This perspective creates two fundamental measurement questions for
marketers: How frequently does a brand enter the generated set, and how
prominently does it appear when recommended? We adapt metrics from
information retrieval and recommender systems to capture these
dimensions. \emph{Brand Recommendation Probability} (BRP@$k$)
measures \emph{prevalence}, or the probability that a brand appears
among up to $k$ recommendations. \emph{Mean Reciprocal Rank}
(MRR@$k$) measures \emph{prominence} by considering both whether
the brand appears and its position in the recommendation list. Because
LLM responses can vary across repeated requests, both quantities are
estimated from replicated queries. Formal definitions and estimation
procedures are provided in the Framework subsection below.

Category-only prompts provide a useful baseline because they specify
what the consumer is seeking but provide no information about goals,
preferences, or constraints. The resulting generated sets reveal which
brands LLMs tend to recommend for the category and which remain absent.
Prior research on conventional recommender systems documents popularity
bias, whereby already popular items receive disproportionate exposure,
while long-tail alternatives are recommended less frequently \cite{fleder2009blockbuster,abdollahpouri2019managing}. Whether LLM-generated sets
similarly favor popular brands or exhibit different patterns of
recommendation prevalence and prominence is an empirical question. We
therefore ask,

RQ1: How are LLM brand recommendations distributed across competing
brands within a product category?

\subsection{Consumer Needs and Brand Positioning}

Category membership alone does not determine consideration.
Consumers' goals and usage situations can change which
alternatives they retrieve \cite{ratneshwar1991substitution, shocker1991consideration,lee2006shopping}. At the same time, firms invest in positioning brands around particular consumer needs, use cases, and
value propositions. Positioning establishes the category as a frame of
reference and then differentiates the brand through meaningful points of
difference intended to make it relevant to particular consumers and
situations \cite{keller2019strategic}. Advertising and other
marketing communications seek to build and reinforce these associations.

LLM-mediated recommendation introduces a new intermediary into this
matching process. Rather than retrieving brands from memory themselves,
consumers can express their needs to an LLM and allow the model to
identify appropriate alternatives. A category-only prompt gives the LLM
little information with which to distinguish among brands positioned for
different needs. A \emph{needs-based prompt}, by contrast, supplies
information about the consumer\textquotesingle s goals, constraints,
preferences, or intended use. Providing such context may therefore
change both which brands enter the generated set and where they are
ranked. This leads to our second research question,

RQ2: How do needs-based prompts shift LLM recommendations toward brands
that fit consumers\textquotesingle{} stated needs?

Finally, brand recommendations may also be associated with the broader 
marketplace information environment. Brands vary substantially in their
advertising presence, news coverage, online conversation, search
interest, and information seeking. These indicators are themselves
highly interrelated and may reflect a broader dimension of marketplace
visibility. Because our observational data cannot establish whether
these signals influence LLM recommendations, reflect common underlying
factors such as consumer interest, or arise through other mechanisms, we
treat their relationship with recommendation prominence as exploratory
rather than causal. We therefore examine the marketplace correlates of
MRR without proposing a directional hypothesis or additional research
question.

\subsection{Framework for Evaluating LLM Brand Recommendations}

\begin{figure}[htb]
\includegraphics[height=4in]{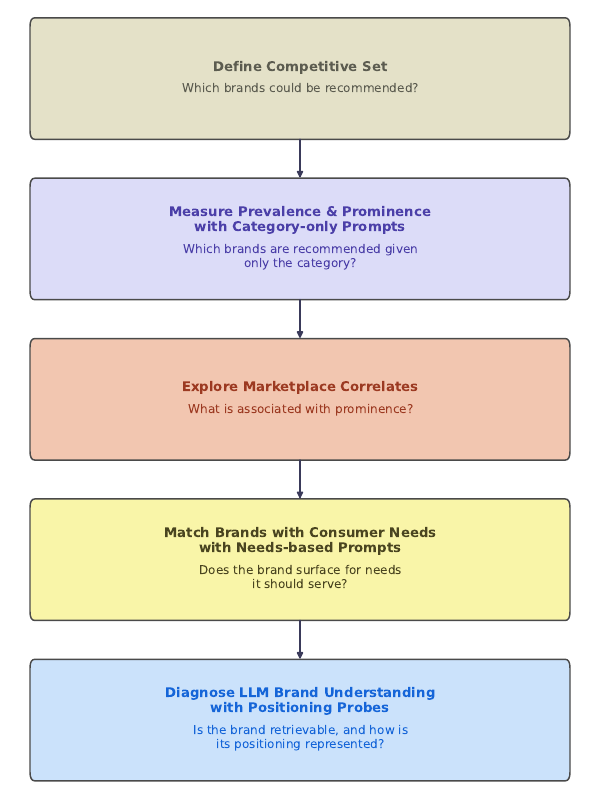}
\caption{A five-step framework for auditing LLM brand recommendations}
\Description{}
\label{fig:fig1}
\end{figure}

Building on the generated-set perspective, we propose a five-step
framework for evaluating LLM brand recommendations (Figure~\ref{fig:fig1}). The
framework can be applied to a focal brand by managers conducting a brand
audit or to an entire category by researchers and market analysts.

\textbf{Define.} The first step is to establish the competitive set
independently of the LLM responses. This is essential because defining
the set from observed recommendations would systematically exclude
brands that LLMs rarely or never recommend, creating a selection bias in
subsequent analyses (especially Explore below). An externally defined
competitive set makes such omissions observable, provides an appropriate
denominator for category-level analysis, and allows marketers to compare
recommended brands with meaningful competitors that fail to enter the
generated set.

\textbf{Measure}. The second step uses \emph{category-only prompts} to
establish baseline brand recommendations, quantifying recommendation
prevalence with Brand Recommendation Probability (BRP@$k$) and
recommendation prominence with Mean Reciprocal Rank (MRR@$k$).
Unlike deterministic rankings, LLMs do not necessarily return the same
brands or ordering when an identical prompt is issued repeatedly.
Recommendation, therefore, should be treated as a probability
distribution rather than a single observed list. We recommend a
two-stage sampling design in which a sample of commercially available
LLMs is selected, and the same prompt is then issued repeatedly to each
LLM in independent sessions. BRP@$k$ for a focal brand is estimated
as the proportion of replicated recommendation lists in which the brand
appears among the first $k$ recommendations. Estimates can be
calculated for individual LLMs or aggregated across models to
characterize recommendation prevalence more broadly. BRP@$k$ is
related to the precision@$k$ metric  \cite{manning2008xml} from
information retrieval. Whereas Precision@$k$ evaluates the
\emph{relevance to the user} of retrieved items for a query,
BRP@$k$ measures the probability that a focal brand appears in a
recommendation list of up to $k$ brands.

MRR@$k$ additionally incorporates the rank of the recommended
position (Voorhees 1999). For focal brand $b$, let $r_{bj}$
denote its rank in replicate $j$. Let the \emph{reciprocal rank} be
$\mathrm{RR}_{bj} = 1/r_{bj}$ when the brand appears in the recommendation
and zero when it is not recommended. Thus, a brand receives a score of 1
when ranked first, 1/2 when ranked second, 1/3 when ranked third, and 0
when absent. Across $n$ replicated queries, brand-specific MRR is
$\mathrm{MRR}_b = \frac{1}{n}\sum_{j = 1}^{n}\mathrm{RR}_{bj}$. MRR, therefore, incorporates both the frequency with which a brand is recommended and its prominence within the list.

\textbf{Explore}. Once differences in recommendation prominence have
been established, external marketplace data can be used to explore
characteristics associated with those differences. Such analyses can
incorporate paid advertising, earned media, online conversation, search
behavior, or other indicators relevant to the category. Identifying
these associations can help managers generate hypotheses about which
marketplace activities and signals may contribute to LLM recommendation
prominence, informing subsequent experimentation, communication
strategies, and potentially resource allocation. Because these
relationships are observational, however, they should guide testing
rather than be interpreted as causal evidence for reallocating marketing
expenditures.

\textbf{Match}. Category-level prominence alone does not establish
whether an LLM recommends appropriate brands for consumers of a
particular focal brand. The sampling procedure used in the Measure stage
can therefore be repeated using needs-based prompts that articulate
goals, constraints, preferences, and intended uses for a target consumer
of the focal brand. BRP@$k$ and MRR@$k$ are calculated in the
same manner, allowing recommendation prevalence and prominence under
specific consumer needs to be compared with the category-only baseline.
This analysis reveals whether the focal brand surfaces when its target
consumers express needs consistent with its intended positioning.

\textbf{Diagnose}. When a brand fails to surface for consumer needs it
is intended to serve, diagnostic probes can be used to investigate the
LLM\textquotesingle s representation of the brand and its positioning. A
first diagnostic is \emph{conditional retrievability}: positioning
probes containing distinctive cues from the brand\textquotesingle s
intended positioning test whether the brand can be surfaced when those
associations are made explicit. More broadly, probes can elicit the
needs, use cases, attributes, points of difference, and competitive
associations that the LLM associates with the brand and compare them
with the firm\textquotesingle s intended positioning. Such diagnostics
can distinguish between a brand that is known to the LLM but not readily
activated by realistic consumer cues and one whose LLM representation
appears incomplete or inconsistent with its intended positioning.

\section{Methodology}

To illustrate the proposed measurement framework, we selected five
product and service categories that represent realistic consumer
decisions for which LLM recommendations are likely to be sought. The
categories were selected using four criteria. First, they represent
relatively high-involvement purchase decisions where consumers are
likely to seek advice rather than make an impulse purchase \cite{mittal1989measuring}. Second, they span a diverse set of products and services to
demonstrate the framework\textquotesingle s generality. Reliable
external benchmark data were available describing the major competing
brands within each category. Finally, each category contains a
sufficiently rich competitive set to illustrate meaningful differences
in recommendation probabilities across brands.

Applying the four criteria, we selected five categories: power tools
(cordless drills), boat cruises, pet food (cat food), home appliances
(coffee makers), and apparel (hiking jackets). Although the BrandZ
benchmark data are organized at the broader category level (e.g., power
tools), consumers typically seek recommendations for more specific
purchase decisions (e.g., asking for a ``cordless drill'' or ``coffee
maker'' versus a ``power tool'' or ``home appliance''). We therefore
designed prompts around the specific products or services that consumers
would naturally request from an LLM while retaining the broader category
definitions to obtain marketplace brand data. Within each category, we
included the major competing brands identified by BrandZ and market
share data from Statista, along with additional brands that appeared in
preliminary LLM recommendation trials. This ensured that the evaluation
captured both established market leaders and brands that LLMs
recommended despite not appearing in the benchmark data.

External marketplace data were assembled to understand differences in
recommendation prominence across brands. Kantar BrandZ provides measures
of brand equity (Salient, Meaningful, and Different) for leading brands
in each category and was used to identify the competitive set. These
data were supplemented with indicators of brands' broader marketplace
visibility: advertising expenditures from Kantar capture paid media;
press mentions from LexisNexis capture earned media; Brandwatch captures
online brand conversation; Google Trends indices capture consumer search
interest; and Wikipedia page views capture brand-related information
seeking. Market-share data from Statista were also used to identify
important competitors not covered by BrandZ. Advertising data were
measured over a rolling 12-month period ending in May 2026; the other
visibility measures covered either calendar year 2025 or a comparable
12-month period. Full definitions, sources, and collection periods are
provided in the supplement.

Because several sources required additional measurement decisions, we
briefly describe their construction. LexisNexis press mentions and
Google Trends were collected using both brand-only and
brand-plus-category searches; we use the latter to reduce ambiguity for
brand names that may occur in unrelated contexts. Because Google Trends
normalizes results within each query, a common anchor brand was included
across query batches within each category, and scores were re-scaled
relative to that anchor, making brands within a category comparable;
values below Google Trends' reporting threshold were coded as zero.
Brandwatch mention volumes were obtained from its historical-data
facility, which estimates total volume from a 5\% random sample of
matching mentions. We therefore interpret Brandwatch values as relative
indicators of online brand conversation rather than exact mention
counts. Additional details, including search specifications, category
anchors, and data-source coverage, are provided in the supplement.

\subsection{Experiment}

We evaluated six commercially available LLMs that consumers commonly use
to obtain product recommendations: OpenAI GPT-5.5 and GPT-5.4 Mini,
Google Gemini 3.1 Pro Preview and Gemini 2.5 Flash, and Anthropic Claude
Opus 4.7 and Claude Sonnet 4.6. The models were accessed
programmatically through their respective provider APIs rather than
through consumer-facing chat interfaces. Each recommendation request was
submitted as a new, stateless API call with no conversation history or
user information. Web search and other external retrieval tools were
disabled, ensuring that each observation represented an independent
response to the prompt.

The first stage used \emph{category-only prompts} to measure baseline
brand recommendations when no information about the consumer or intended
use was provided. The standardized prompt was: ``I am looking for a
{[}category{]}. Return valid JSON only, with this schema:
\{``brands'':{[}``Brand name''{]}\}. Include up to five brand names. Do
not include explanations, markdown, numbering, or any text outside the
JSON object.'' The category placeholder was replaced with cordless
drill, hiking jacket, coffee maker, cat food, or boat cruise; all other
wording was held constant. For each of the five categories and six
models, the prompt was issued 40 times in independent sessions,
producing 1,200 recommendation lists.

The second stage used detailed needs-based prompts to examine how
recommendations differed when the LLM received information about
consumers' goals, constraints, preferences, and intended uses rather
than only a product category. For the analyses reported here, we used
100 detailed prompts across the five categories, with 20 consumer
scenarios per category. The prompts varied in budget, experience,
intended use, desired performance, and other purchase considerations.
All prompts requested up to five brands using the same standardized JSON
format as the category-only prompts. The prompt library and condition
labels are available in the online supplement.

Each detailed needs-based prompt was submitted to all six LLMs twice,
with each repetition conducted as an independent stateless request. This
produced 1,200 recommendation lists (100 prompts $\times$ 6 models $\times$ 2
repetitions). We use these responses to examine how consumer context
changes recommendation prevalence and prominence relative to the
category-only baseline.

Model responses were preserved in their original form and then processed
to extract brand names in the recommended order. We retained the first
five unique brands and standardized names using category-specific alias
dictionaries that accounted for spelling and formatting variants,
abbreviations, parent-brand references, and product-line names. When
multiple recommendations mapped to the same brand, only its
highest-ranked occurrence was retained. Brand-level metrics were then
calculated separately for each combination of product category, model,
prompt condition, and brand.

\section{Results}

\subsection{Measuring Brand Recommendation Prevalence and Prominence}

\begin{figure*}[htb]
\includegraphics[height=5in]{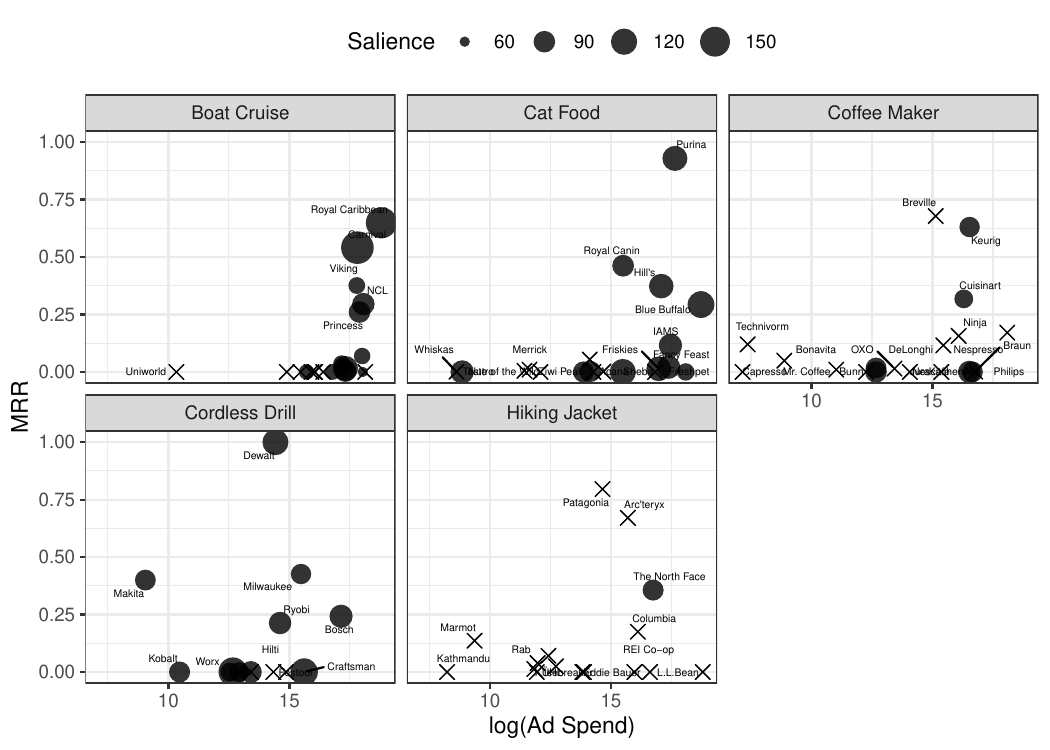}
\caption{Relationship between LLM recommendation prominence (MRR@5) and advertising expenditures, by product category}
\Description{}
\label{fig:mrr}
\end{figure*}

\begin{figure*}[htb]
\includegraphics[height=5in]{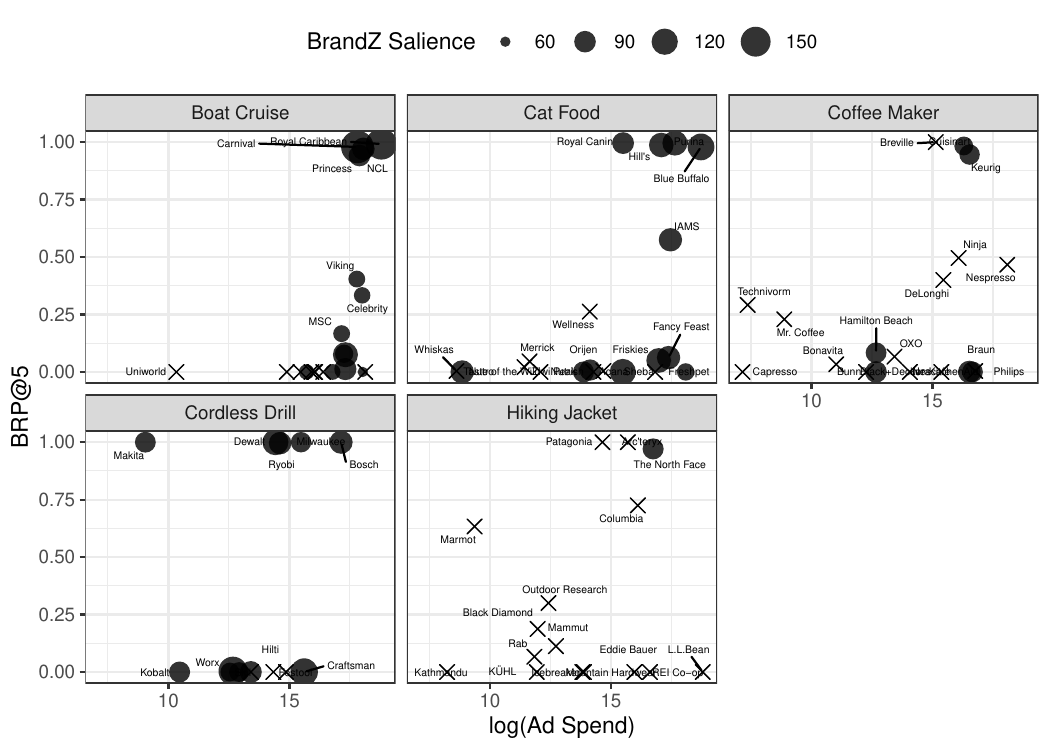}
\caption{Relationship between LLM recommendation prevalence (BRP@5) and advertising expenditures, by product category}
\Description{}
\label{fig:brp}
\end{figure*}

We first consider the category-only prompts. Figures~\ref{fig:mrr} and \ref{fig:brp} plot MRR and
BRP@5, respectively, against the log of advertising expenditures, with
separate panels for each category. Advertising expenditure provides a
widely available measure of brands\textquotesingle{} paid marketplace
presence and is available for most brands in our sample. BrandZ salience
is represented by the size of the circular plotting symbols. Salience is
available only for brands tracked by BrandZ, which tend to be the
leading brands in each category; brands not tracked by BrandZ are
represented by an X. The complete data underlying these analyses and
figures are available on our anonymous GitHub site,\footnote{
\href{https://anonymous.4open.science/r/LLM-Monitor-CF9F}{\ul{https://anonymous.4open.science/r/LLM-Monitor-CF9F}}}
allowing
readers to examine results for individual brands and categories in
greater detail.

Three findings stand out. First, many large, established brands receive
no recommendations at all. These include L.L.Bean, Eddie Bauer, and REI
Co-op in hiking jackets; Craftsman and Black+Decker in cordless drills;
Braun and Philips in coffee makers; Freshpet in cat food; and Virgin
Voyages and Regent Seven Seas in cruises. Thus, even well-established
brands can be effectively absent from LLM-generated consideration sets.
From a managerial perspective, such omissions are consequential: brand
managers should monitor not only how prominently their brands are
recommended, but whether they are recommended at all.

Second, the figures provide only limited evidence of the popularity bias
documented in the marketing and recommender-systems literature. We
operationalize brand popularity using BrandZ salience, represented by
plotting symbol size. The clearest evidence occurs in cruises, where
Carnival (salience=177) and Royal Caribbean (162), the two most salient
brands, also have among the highest MRR values. Yet even in this
category, the relationship is inconsistent: Viking receives substantial
recommendation prominence despite much lower salience (70), while highly
salient Disney Cruise Line (111) is almost never recommended (MRR=.015).
Across the remaining categories, salience and recommendation prominence
show little systematic relationship. In some categories, however, a
different pattern emerges. For cordless drills and hiking jackets, LLM
recommendations tend to favor higher-end or premium brands, such as
DeWalt, Milwaukee, Patagonia, and Arc'teryx, while giving relatively
little recommendation prominence to more mass-market brands such as
Black+Decker, Craftsman, Eddie Bauer, and L.L.Bean. This pattern,
however, does not hold across all categories. Thus, the familiar
popularity-bias explanation provides, at best, a partial account of
which brands LLMs recommend; in some categories, brand positioning or
market tier appears to provide an alternative explanation.

Third, BRP@5 and MRR generally tell a consistent story, but MRR provides
additional discrimination when brands have similar recommendation
prevalence. For example, DeWalt, Milwaukee, Makita, Bosch, and Ryobi all
have BRP@5=1 for cordless drills, indicating that each appears in every
recommendation set. MRR, however, reveals substantial differences in
prominence: DeWalt is consistently ranked highest, followed by Milwaukee
and Makita, while Bosch and Ryobi rank lower. A similar pattern occurs
for cruises, where Royal Caribbean, Carnival, Viking, Norwegian Cruise
Line, and Princess all have BRP@5=1, but MRR distinguishes their
positions within the recommendation set. Cat food provides yet another
example: Purina, Royal Canin, Hill's, Blue Buffalo, and IAMS all have
BRP@5=1, yet MRR differentiates their recommendation prominence. Thus,
BRP provides an intuitive measure of whether brands consistently enter
the LLM-generated consideration set, whereas MRR reveals competitive
differences that prevalence alone can obscure.

\subsection{Explaining Brand Recommendation Prominence}

\begin{figure}
\includegraphics[height=3.5in]{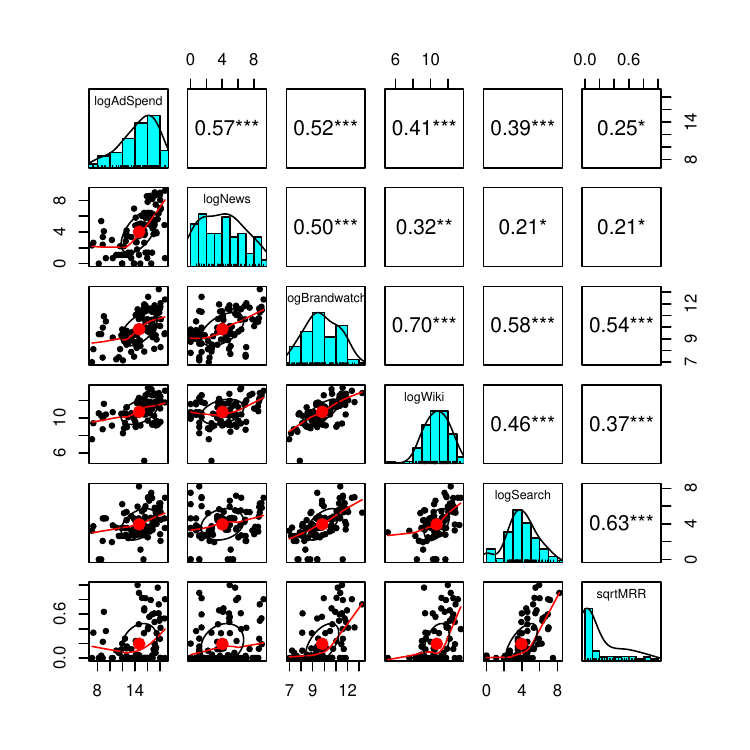}
\caption{Relationships among marketplace visibility measures and LLM recommendation prominence}
\Description{}
\label{fig:splom}
\end{figure}

If popularity does not explain recommendations, a key question is: what
does? We use the five marketplace-visibility measures described
previously (advertising expenditures, press mentions, online brand
conversation, search interest, and Wikipedia page views) to explore
their associations with recommendation prominence. Figure~\ref{fig:splom} shows a
scatterplot matrix of the log-transformed visibility measures and the
square root of MRR, with Pearson correlations reported in the upper
triangle. Transformations were used to reduce skewness and the influence
of outliers. The scatterplot reveals two statistical issues that will
complicate modeling. First, there are substantial correlations between
many of the explanatory variables, indicating multicollinearity. For
example, the correlation between Brandwatch and Wikipedia is .70,
suggesting that these variables can roughly serve as surrogates for one
another in a model. The eigenvalues of the correlation matrix of
predictors are 2.9, 0.90, 0.56, 0.41, and 0.24; the dominant first
dimension suggests that the five measures share a substantial common
component that we interpret as \emph{marketplace visibility}.
Multicollinearity will destabilize our models by inflating coefficient
standard errors, causing the magnitude of the coefficients to be highly
sensitive to which other variables are included in the model, and
possibly even flipping signs of some coefficients \cite{kutner2005applied}.
Second, there are many 0 values of the dependent variable, which is
bounded between 0 and 1. Normality assumptions of linear models will not
hold and some predictions could be outside the {[}0,1{]} range. The
fractional logit model \cite{papke1996econometric} is a reasonable way
to address the second issue and can be estimated with R's glm and glmnet
functions, specifying the quasibinomial family.

There is a third issue that is conceptual rather than statistical. We
are not estimating a confirmatory model and make no causal claims about
the relationships that emerge. The analysis should instead be
interpreted as an exploratory observational study \cite[p. 345]{kutner2005applied}. For example, consider cor(Google Trends, sqrtMRR) = .63.
This association does not imply that greater search activity causes
greater LLM recommendation. Both could reflect unobserved consumer
interest, brand relevance, or other marketplace activity. Moreover, the
causal ordering among the marketplace measures themselves is unclear:
online conversation may stimulate search, search may stimulate
conversation, or both may respond to common events. The scatterplot
matrix shows that all five marketplace measures are significantly and
positively correlated with MRR. These relationships should be
interpreted as exploratory associations that require further
theorization and testing.

Returning to the multicollinearity issue, ridge regression has a long
history in the marketing literature as a means of stabilizing estimates when predictors are highly correlated  \cite{mahajan1977parameter, malthouse1999ridge},
including settings such as marketing-mix models where marketing
activities commonly move together. It shrinks coefficients toward zero,
trading some bias for reduced variance and greater stability under
multicollinearity. The amount of shrinkage is governed by hyperparameter
($\lambda$): $\lambda$=0 corresponds to the unpenalized model, with coefficients
increasingly shrunk toward zero as $\lambda$ increases. Because all five
marketplace measures are theoretically plausible predictors and share a
substantial common dimension, we use ridge regression and select $\lambda$ using
cross-validation, following \cite[p.\ 246]{james2021introduction}. We also
estimate lasso models as a robustness check; unlike ridge, lasso can shrink coefficients exactly to zero and therefore performs variable selection.

\begin{figure}[htb]
\includegraphics[width=3.5in,height=3in]{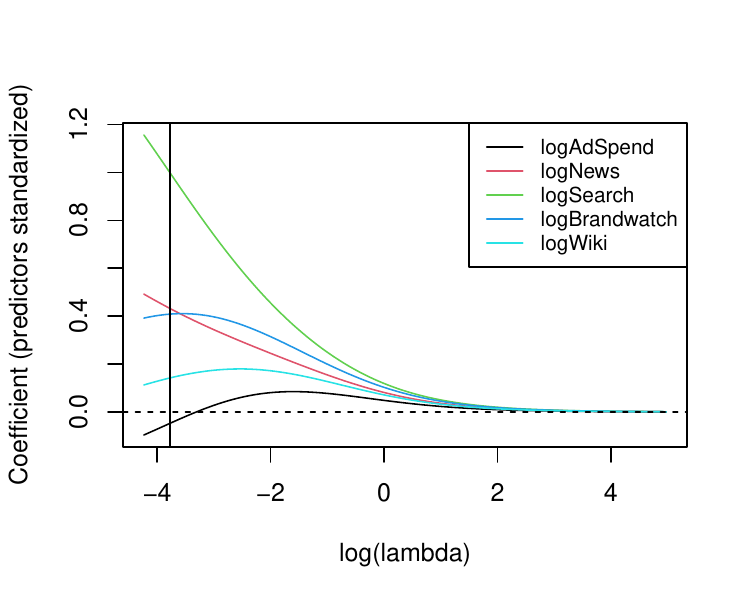}
\caption{Ridge trace for marketplace visibility predictors of
LLM recommendation prominence}
\label{fig:ridge}
\Description{Lines show standardized coefficient estimates across values
of the ridge penalty parameter ($\lambda$); vertical red line indicates the
value of $\lambda$ selected by cross-validation (1SE).}
\end{figure}

We regressed MRR on the five marketplace measures and category dummy
controls using the fractional logit with ridge regularization
($n=82$). Because the marketplace measures are expressed in
different units, we standardized the five predictors before estimation,
allowing their coefficient magnitudes to be compared. Figure~\ref{fig:ridge} presents
the ridge trace, plotting the coefficients of the five marketplace
predictors against the shrinkage parameter $\lambda$ on a log scale. The
vertical red line at $\log(\lambda)=-2.39$ indicates the value selected by
cross-validation using the 1SE criterion. At this value, logSearch has
the largest coefficient, indicating the strongest predictive
relationship with MRR. The coefficients for logBrandwatch is the next
largest, whereas the coefficients for logNews, logWiki and logAdSpend
are close to zero. This ordering is robust across alternative model
specifications. In particular, logSearch remains the strongest predictor
in fractional logit models without shrinkage (Table~\ref{tab:logit}), while a cross-validated lasso model retains logSearch and logBrandwatch but shrinks logNews, logWiki and logAdSpend to zero using cross-validation. As discussed earlier, these results should nevertheless be interpreted as exploratory predictive relationships rather than causal effects.

\begin{table}[!htbp] \centering
  \caption{Fractional logit regression results}
\label{tab:logit}
\centering
\begin{tabular}{lrrrr}
  \hline
 & Estimate & Std.\ Error & $t$ value & Pr$(>|t|)$ \\ 
  \hline
Intercept              &$-12.2898$& 2.6596 &$-4.62$& 0.0000 \\ 
Category               &          &        &       & \\
  \quad Boat cruise    & 0        &        &       & \\
  \quad Cat Food       & 3.9459   & 1.0025 & 3.94  & $0.0002^{***}$ \\ 
  \quad Coffee maker   & 1.6462   & 0.8433 & 1.95  & 0.0548 \\ 
  \quad Cordless drill & 3.6836   & 1.2943 & 2.85  & $0.0058^{***}$ \\ 
  \quad Hiking jacket  & 3.7494   & 1.9641 & 1.91  & 0.0603 \\ 
  logAdSpend           &$-0.1434$ & 0.1040 &$-1.38$& 0.1723 \\ 
  logWiki              &$-0.1290$ & 0.2434 &$-0.53$& 0.5978 \\ 
  logNews              & 0.5844   & 0.2673 & 2.19  & $0.0321^{*}$\\ 
  logSearch            & 1.0586   & 0.2149 & 4.93  & $0.0000^{***}$ \\ 
  logBrandwatch        & 0.3385   & 0.2417 & 1.40  & 0.1658 \\ 
   \hline
\end{tabular}
\end{table}

\subsection{Needs-based Prompts}

We next illustrate how the framework can be extended to needs-based
prompts using two brands that were never recommended in response to
category-only prompts: Craftsman drills and L.L.Bean hiking jackets.
Rather than asking only for brands in a product category, needs-based
prompts describe a plausible consumer, use case, goals, and constraints,
allowing the LLM to match those needs against its knowledge of competing
brands and their positioning. These analyses are illustrative rather
than a systematic comparison across all brands because they focus on two
selected brands that were omitted under category-only prompting.

Craftsman is positioned primarily toward DIY consumers seeking
dependable, accessible tools for home improvement and everyday projects.
L.L.Bean primarily serves consumers seeking practical, durable, and
versatile outdoor products for everyday recreation and family activities
rather than exclusively technical or high-performance outdoor pursuits.
For the Craftsman illustration, we used the 20 cordless-drill prompts
from the needs-based prompt library described in Methods. For the
L.L.Bean illustration, we used the 20 hiking-jacket prompts. These
prompts described realistic consumers, use cases, goals, and constraints
without mentioning Craftsman or L.L.Bean or supplying their exact
positioning. Each prompt was submitted to all six LLMs twice in
independent stateless requests, producing 240 recommendation lists for
each focal brand.

For illustration, one cordless-drill prompt asked: ``I am the kind of
person who watches YouTube tutorials and then spends the weekend
building things. I need a cordless drill that can keep up, good enough
for most home projects, but I am not a professional contractor.'' One
detailed hiking-jacket prompt stated: ``I am a teacher and I take my
students on outdoor field trips. I need a waterproof hiking jacket that
is tough, affordable, and available in normal sizes. Something
practical, not trendy.'' All prompts used the same response instructions
as the category-only prompts, requesting up to five brand names in a
standardized JSON format.

Neither Craftsman nor L.L.Bean appeared in the category-only
recommendations (MRR = 0; BRP@5 = 0\%). Under the detailed needs-based
prompts, Craftsman achieved an MRR = .108 and a BRP@5 = 35.4\%, whereas
L.L.Bean achieved an MRR = .025 and a BRP@5 = 5.4\%. Each estimate is
based on 240 recommendation lists. These results show that detailed
information about the consumer and intended use can bring previously
omitted brands into the LLM-generated recommendation set, but the effect
remains limited. Craftsman appeared in slightly more than one-third of
the recommendation lists and was generally not ranked prominently, while
L.L.Bean remained rarely recommended. From a managerial perspective, the
results suggest that realistic, context-rich consumer requests do not
consistently retrieve either brand, particularly L.L.Bean.

These results raise an important question: are Craftsman and L.L.Bean
omitted because the LLMs lack knowledge of their positioning, or because
relevant brand knowledge is not activated by ordinary needs-based
prompts? To investigate these possibilities, we created more
\emph{diagnostic positioning probes} that incorporated language drawn
directly from the brands' own marketing materials. We call them probes
rather than prompts because they are intentionally unrealistic as
consumer requests: a typical shopper seeking advice about a cordless
drill or hiking jacket would be unlikely to reproduce detailed
brand-positioning language without first researching the market. When
supplied with these diagnostic cues, however, BRP@5 increased to 81.3\%
for Craftsman and 88.5\% for L.L.Bean. These results demonstrate that
the brands are retrievable when the LLMs are given sufficiently
diagnostic cues consistent with their intended positioning. They
therefore suggest that low recommendation rates under category-only and
needs-based prompts cannot be attributed simply to an absence of brand
knowledge. At the same time, conditional retrievability does not
establish that the LLMs hold complete or accurate representations of
either brand's positioning. Additional probes could systematically test
whether the attributes, points of difference, target users, and use
cases that LLMs associate with a brand correspond to its intended
positioning.

\section{Discussion}

Our empirical analyses reveal several patterns in LLM brand
recommendations. First, category-only prompts across six LLMs frequently
omit large, established brands entirely, indicating that conventional
marketplace presence does not guarantee inclusion in LLM-generated
recommendation sets. Second, we find very limited evidence of the
popularity bias documented for conventional recommender systems: salient
brands are sometimes recommended prominently, but the relationship
varies substantially across categories. In cordless drills and hiking
jackets, recommendations instead tend to favor higher-end or premium
brands while neglecting more mass-market alternatives, although this
pattern does not generalize across categories. Third, recommendation
prominence is systematically associated with observable marketplace
signals. Google search interest provides the strongest and most robust
predictive signal, followed by online brand conversation, while news
mentions, advertising expenditures and Wikipedia page views contribute
relatively little once the correlated measures are considered jointly.
These relationships are exploratory and should not be interpreted
causally. Finally, our needs-based examples demonstrate that providing
information about consumers' goals and constraints can cause omitted
brands to enter the recommendation set, while more diagnostic
positioning probes show that low recommendation rates do not necessarily
reflect a lack of LLM knowledge about the brand. Taken together, the
findings suggest that LLM brand recommendation reflects both the
visibility of brands in the marketplace and the extent to which
available prompt information activates brand associations relevant to
the consumer's needs.

\subsection{Theoretical Contributions}

This study extends choice-architecture and consideration-set theory to
LLM-mediated decisions. Traditional choice architectures operate on an
identifiable set of alternatives: retailers stock an assortment, search
engines rank existing pages, and recommender systems select and rank
alternatives from an explicit candidate set. LLMs differ because the
brands eligible to appear are not directly specified or observable, and
repeated responses to the same prompt may produce different sets. We
call the resulting alternatives a \emph{generated set}. This extends
consideration-set theory by moving brand retrieval from the
consumer\textquotesingle s memory to an external, stochastic
intermediary. Consequently, the relevant question is not only how
prominently a brand appears but whether it enters the generated set at
all. BRP captures the prevalence of this inclusion, while MRR captures
recommendation prominence; because both can vary across repeated
queries, they must be estimated through sampling rather than read from a
fixed recommendation list.

Second, our findings qualify the applicability of conventional
popularity-bias explanations to LLM brand recommendations.\linebreak Recommender-systems research has shown that popular alternatives can
receive disproportionate recommendation exposure \cite{fleder2009blockbuster,abdollahpouri2019managing,abdollahpouri2021user}. We find only limited
correspondence between BrandZ salience and LLM recommendation
prominence, with substantial variation across categories. Instead,
recommendation prominence is associated with a broader
marketplace-visibility environment, particularly consumer search
interest, news coverage, and online brand conversation. Because these
measures are highly correlated and observational, we cannot identify the
mechanisms producing these relationships. The findings nevertheless
suggest that marketplace popularity alone is insufficient to explain
which brands LLMs surface and point to marketplace visibility and brand
positioning as important areas for further theory development.

Third, the needs-based and diagnostic results extend branding theory
into LLM-mediated recommendation. Brand positioning seeks to establish
associations between a brand and particular consumer needs, use cases,
and points of difference \cite{keller2019strategic}. Our results show
that brands omitted from category-only recommendations can become more
prevalent when consumers articulate relevant needs, making the LLM an
intermediary in matching consumer needs with brand positioning.
Moreover, positioning probes demonstrate that omission from ordinary
recommendations does not necessarily imply that a brand is unavailable
to the model: Craftsman and L.L.Bean became highly retrievable when
distinctive positioning cues were supplied. This distinction between
\emph{recommendation} and \emph{conditional retrievability} suggests a
new branding question: not simply whether brand information is
represented by an LLM, but whether that representation is activated when
consumers express needs the brand is intended to serve.

\subsection{Practical Implications}

Our findings suggest two broad implications for brand managers. First,
managers should consider the marketplace information environment
surrounding their brands. Recommendation prominence is positively
associated with a common marketplace-visibility dimension reflected in
advertising, search, news coverage, online conversation, and information
seeking. However, not all manifestations of visibility are equally
predictive. The ridge analysis identifies Google search interest as the
strongest signal, followed by online brand conversation. These associations should
not be interpreted causally or as evidence that managers should attempt
to optimize Google Trends or Brandwatch measures directly. Rather, they
suggest investigating which marketing activities generate sustained
marketplace attention, particularly search and online conversation, and
whether increases in such attention are subsequently associated with
greater LLM recommendation prominence.

Second, visibility alone is unlikely to be sufficient; what the brand is
associated with also matters. Our findings therefore reinforce a
fundamental principle of brand positioning: brands should establish the
points of parity necessary for category membership while developing a
small number of distinctive and relevant points of difference \cite{keller2019strategic}. In an LLM-mediated environment, consistency in these
associations may become especially important because brand information
is represented across firms\textquotesingle{} own communications and a
broader ecosystem of news, reviews, online conversations, and other
digital content. Managers should seek to communicate the intended
positioning consistently across the touchpoints they control and
cultivate congruent associations across those they can influence. The
objective is not simply to maximize the volume of information about the
brand, but to establish a coherent set of associations between the brand
and the consumer needs for which it should be considered. When consumers
articulate those needs to an LLM, a clearly and consistently positioned
brand may be easier for the LLM to match to them.

\subsection{Limitations and Future Research}

Our illustrative needs-based and diagnostic-probe results for Craftsman
and L.L.Bean come from a small set of handwritten prompts built
specifically for these two brands, separate from the systematic
100-prompt needs-based study reported elsewhere. This three-stage
approach needs validation at scale across a larger, randomly selected
set of brands before it\textquotesingle s a general diagnostic tool. The
price-tier pattern held in only two of five categories, and we
don\textquotesingle t yet know what distinguishes those from the other
three. Future work should extend the probe design more broadly by
varying the attributes, target users, use cases, and points of
difference associated with a focal brand, and then comparing those
associations with the brand's intended positioning. This would allow
researchers to distinguish between two problems: the LLM may know a
brand but fail to recommend it, or it may hold an incomplete or
inaccurate representation of what the brand stands for. A related
research direction is to develop and test methods to improve that
representation by supplying LLMs with authoritative positioning
information and examining whether the resulting brand associations and
recommendations are more consistent with the firm's intended
positioning. Our study used fresh LLM sessions and therefore did not
incorporate users' prior chat histories. In practice, consumers may have
extensive interaction histories that provide LLMs with information about
their preferences, interests, and prior decisions, thereby influencing
brand recommendations. Future research could examine whether
category-specific prompts issued by a sample of consumers, using their
own LLM accounts and prompt histories, produce different recommendation
patterns than those observed in fresh sessions.
\bibliographystyle{ACM-Reference-Format}
\bibliography{mybibliography.bib}

\newpage
\appendix

\section{Additional Data Details}
\label{app:data}

\begin{table*}[t]
\caption{Data Sources, Collection Windows, and Per-Category Details}
\label{tab:data-sources}
\centering
\small
\begin{tabularx}{\textwidth}{
    @{}
    >{\raggedright\arraybackslash}p{3.0cm}
    >{\raggedright\arraybackslash}p{2.2cm}
    >{\raggedright\arraybackslash}p{2.1cm}
    >{\raggedright\arraybackslash}p{1.6cm}
    >{\raggedright\arraybackslash}X
    @{}
}
\toprule
\textbf{Measure} &
\textbf{Source} &
\textbf{Window} &
\textbf{Retrieved} &
\textbf{Per-category notes} \\
\midrule
Meaningful / Different / Salient / Demand power / Pricing power &
Kantar BrandZ &
Most recent wave per category &
--- &
Cordless Drills Sep 2022; Boat Cruise May 2023; Coffee Maker Apr 2024;
Hiking Jacket May 2025; Cat Food Jul 2025 \\

\addlinespace
Brand spend / Ad spend & Kantar & Rolling 12 months &
May 18--19, 2026 & Uniform across categories \\

\addlinespace
Wikipedia pageviews & Wikipedia Pageviews Analysis &
Jan 1--Dec 31, 2025 & --- & Uniform across categories \\

\addlinespace
News, brand-only & Nexis Uni &
Jan 1--Dec 31, 2025 & --- & Query = ``[brand]'' \\

\addlinespace
News, brand + category & Nexis Uni &
Jan 1--Dec 31, 2025 & --- &
Cruise AND ``cruise''; Cat Food AND ``cat food'';
Coffee AND (``coffee maker'' OR ``coffee machine'');
Drills AND (``cordless drill'' OR ``power drill'');
Jacket AND ``hiking jacket'' \\

\addlinespace
Google Trends, brand-only &
Google Trends (US, Web) &
Past year &
May 20, 2026 &
Anchors: Cruise = American Cruise Lines; Cat Food = Blue Buffalo;
Coffee = Mr.\ Coffee; Drills = Ryobi; Jacket = Outdoor Research \\

\addlinespace

Google Trends, brand + category &
Google Trends (US, Web) &
Past year &
May 20, 2026 &
Category term appended; Jacket anchor switches to Marmot \\

\addlinespace

Market share &
Statista &
Heterogeneous &
--- &
Coffee Maker survey $n=1{,}199$ (Sep 11--Oct 28, 2025);
Hiking Jacket Nov 2024; Cat Food May 2019;
Boat Cruise survey Jun 24--Jul 20, 2019;
Cordless Drills survey Nov 27--Dec 2, 2017 \\
\bottomrule
\end{tabularx}
\end{table*}

\begin{table*}[t]
\caption{Category Disambiguation Terms Used in the Brandwatch Boolean Queries}
\label{tab:brandwatch-terms}
\centering
\small
\begin{tabularx}{\textwidth}{
    @{}
    >{\raggedright\arraybackslash}p{3.2cm}
    >{\raggedright\arraybackslash}X
    @{}
}
\toprule
\textbf{Product category} &
\textbf{Category disambiguation terms (combined with OR)} \\
\midrule

Boat cruise &
cruise OR cruises OR cruising OR ``cruise line'' OR ``cruise ship'' \\

\addlinespace

Cat food &
``cat food'' OR ``cat treats'' OR kitten OR feline OR kibble OR
``wet food'' OR ``dry food'' \\

\addlinespace

Coffee maker &
``coffee maker'' OR ``coffee machine'' OR espresso OR coffee OR brewer \\

\addlinespace

Cordless drill &
drill OR drills OR ``power tool'' OR ``power tools'' OR cordless OR driver \\

\addlinespace

Hiking jacket &
jacket OR ``rain jacket'' OR hiking OR outdoor OR ``gore-tex'' OR parka \\

\bottomrule
\end{tabularx}

\medskip
\begin{minipage}{\textwidth}
\footnotesize
\textit{Note.} Each query took the form: ``[brand name]'' AND
(terms above). Terms in quotation marks were matched as exact phrases.
\end{minipage}
\end{table*}

\end{document}